\documentclass[a4paper,12pt]{article}
\usepackage[pctex32]{graphics}
\usepackage{amssymb,amsmath}

\begin{document}
\topmargin 0pt
\oddsidemargin 0mm
\newcommand{\be}{\begin{equation}}
\newcommand{\ee}{\end{equation}}
\newcommand{\ba}{\begin{eqnarray}}
\newcommand{\ea}{\end{eqnarray}}
\newcommand{\fr}{\frac}
\renewcommand{\thefootnote}{\fnsymbol{footnote}}

\begin{titlepage}

\vspace{5mm}
\begin{center}
{\Large \bf Classical stability of BTZ black hole \\ in new massive
gravity }

\vskip .6cm
 \centerline{\large
 Yun Soo Myung$^{1,a}$, Yong-Wan Kim $^{1,b}$, Taeyoon Moon $^{2,c}$,
and Young-Jai Park$^{2,3,d}$}

\vskip .6cm

{$^{1}$Institute of Basic Science and School of Computer Aided
Science,
\\Inje University, Gimhae 621-749, Korea \\}

{$^{2}$Center for Quantum Spacetime, Sogang University, Seoul 121-742, Korea \\}

{$^{3}$Department of Physics and Department of Service Systems Management and Engineering, \\
Sogang University, Seoul 121-742, Korea}

\end{center}

\begin{center}

\underline{Abstract}
\end{center}
We study the stability of the BTZ black hole in the new massive
gravity. This  is  a nontrivial task because the linearized
equation around the BTZ black hole background is a fourth order
differential equation.  Away from the critical point of
$m^2\ell^2=1/2$, this fourth order equation is split into two
second order equations: one describes  a massless graviton and the
other is designed for  a massive graviton, which could be obtained
from the Fierz-Pauli action. In this case, calculating quasinormal
modes leads to confirm the stability of the BTZ black hole. At the
critical point, we derive two left and right logarithmic
quasinormal modes from the logarithmic conformal field theory.
Finally, we identify two $s$-massive modes propagating on the
black hole background through the conventional black hole
stability analysis.

\vspace{5mm}

\noindent PACS numbers: 04.70.Bw, 04.60.Kz, 04.70.-s \\
\noindent Keywords: New massive gravity; stability of black hole; quasinormal modes

\vskip 0.8cm

\vspace{15pt} \baselineskip=18pt
\noindent $^a$ysmyung@inje.ac.kr \\
\noindent $^b$ywkim65@gmail.com\\
\noindent $^c$tymoon@sogang.ac.kr\\
\noindent $^d$yjpark@sogang.ac.kr

\thispagestyle{empty}
\end{titlepage}

\newpage
\section{Introduction}

It is well known that Einstein gravity in three dimensions has no
propagating degrees of freedom.   Massive generalizations of
three-dimensional gravity  allow propagating degrees of freedom.
Topologically massive gravity (TMG) is the well-known example
obtained by including  a gravitational Chern--Simons term with
coupling $\mu$~\cite{DJT, DJT2}. The model was extended by the
addition of a cosmological constant term $\Lambda=-1/\ell^2$  to
cosmological topologically massive gravity (CTMG)~\cite{Deser82}.
The gravitational Chern--Simons term is odd under parity and as a
result, the theory shows  a single massive propagating degree of
freedom of a given helicity, whereas the other helicity mode remains
massless. The single massive field is realized as a massive scalar
$\varphi=z^{3/2}h_{zz}$ when using the Poincare coordinates
$x^{\pm}$ and $z$  covering the AdS$_3$ spacetimes~\cite{CDWW}.
However, it was claimed that the negative-energy massive graviton
disappears at the critical point of $\mu\ell=1$~\cite{LSS}. This
cosmological topological massive gravity at the critical point
(CCTMG) may be described by the logarithmic conformal field theory
(LCFT)~\cite{GJ,Myung} even for the zero central charge $c_L=0$.

Another massive generalization of Einstein gravity in three
dimensions was proposed recently by adding a specific quadratic
curvature term to the Einstein-Hilbert action~\cite{bht,bht2}.  This term
was designed to reproduce  the ghost-free Fierz-Pauli action for a
massive propagating graviton in the linearized approximation.
This gravity theory proposed by Bergshoeff, Hohm, and Townsend (BHT)
became known as new massive gravity (NMG). Unlike
the TMG, the NMG preserves parity.  As a result, the gravitons acquire
the same mass for both helicity states, indicating two massive
propagating degrees of freedom. It was shown that there is no ghost
in the linearized BHT gravity by performing a canonical analysis in
flat spacetimes~\cite{Deser}. Furthermore, in flat and de Sitter
spacetimes, the authors~\cite{GST} have found two massive
propagating scalars $\sigma$ and $\phi$ derived from the metric
perturbations, satisfying the Klein-Gordon equation
$[\nabla^2-m^2]\{\sigma,\phi\}=0$. However, up to now, there is no
explicit form of two massive scalars propagating on AdS$_3$
spacetimes, even for the fourth order linearized equation for
graviton was known under the transverse and traceless
gauge~\cite{nmg-3}.

It is well known that the BTZ black hole as solution to Einstein
gravity with $\Lambda$ is also a black hole solution to the CTMG.
However, this does not necessarily imply that there is no difference
in the dynamics of perturbations. It is obvious that perturbation
discriminates between Einstein gravity and CTMG.  Recently, it was
shown that the (non-rotating) BTZ black hole is stable for all
values of $\mu$ against the metric perturbations in the TMG by
considering left-and right-moving normal modes~\cite{Birm}. They
have confirmed the stability by solving the massive scalar equation
of $(\nabla^2_{\rm BTZ}-m^2)\varphi=0$.

In this work, we wish to perform  stability analysis on the  BTZ
black hole in the NMG, which is a nontrivial task.
 The
basic idea of performing the black hole stability is that one
decouples the second order linearized equations and then, manages to
arrive at the Schr\"odinger-type equation with an effective
potential for physically propagating fields~\cite{ReggeW,Chan}. If
all potentials are positive for whole range outside a event
horizon, the black hole under the consideration is stable. If an
effective potential is not positive definite everywhere outside a
horizon, a special trick called the  $S$-deformation technique may
be used to prove the stability~\cite{IK}.  It is well  known that a
practical tool for testing stability of all kinds of black holes is
a numerical investigation of quasinormal frequencies
$\omega=\omega_R-i\omega_I$ by imposing the boundary condition:
ingoing waves near a event horizon and the Dirichlet boundary
condition at infinity~\cite{KZ}. That is, the unstable mode is
defined by the condition of
\begin{equation}
 \omega_R=0,~~\omega_I<0 \label{unstablec}.
\end{equation}
in quasinormal mode approach~\cite{MMS}.  We wish to perform the
stability analysis of the BTZ black holes by computing quasinormal modes of the NMG.
However, it seems that this analysis is
not a straightforward task in the NMG because the linearized
equation is a fourth order differential equation. If the mass
parameter $m^2$ is off the critical value ($m^2\not=1/2\ell^2$), the
fourth order equation split into two second order equations with
transverse and traceless gauge conditions: one is for a massless
graviton (gauge artefact) and the other is for  a massive graviton,
which takes a similar form obtained from the Fierz-Pauli action.
However, at the critical point of $m^2=1/2\ell^2~(c_L=0,c_R=0)$, the
fourth order equation leads to
$[\bar{\nabla}^2-2\Lambda]^2h_{\mu\nu}=0$, which is difficult to be
solved unless the LCFT is introduced. Hence, this case is surely
beyond the standard stability analysis of a black hole prescribed
above. Recently,  Sachs and Solodukhin have determined
quasinormal mode of black hole spectrum for tensor perturbations in
the TMG~\cite{Sachs}. In their operator calculation, they have used the
chiral highest weight condition of $\bar{L}_1h_{\mu\nu}=0$ to derive
quasinormal frequencies, which are similar to the scalar
quasinormal modes. However, this method is inappropriate  to derive
quasinormal modes at the critical point $\mu=1/\ell$ of the TMG. To
this end,  Sachs has generalized the one-to one correspondence
between quasinormal modes in the BTZ black hole and the poles of the
retarded correlators in the boundary conformal field theory to
include logarithmic operators~\cite{Sachs2}. On the other hand, Liu
and Wang have studied the  stability of the BTZ black string against
gravitational perturbations in four dimensions, which is very
similar to the Fierz-Pauli action in three dimensions~\cite{LW}.

The organization of our work is as follows. In section 2, we
briefly review how  the BTZ black hole comes out from the NMG. We
derive quasinormal modes of the NMG by solving the first order
equations with $\Lambda=-1$, which is  based on Sachs and
Solodukhin method used in the TMG  in section 3. However, the
left- and right-moving modes are not orthogonal, which has a
problem to be considered as two independent massive modes. Section
4 is focused on deriving quasinormal modes at the critical point
of $m^2=1/2$ with $\Lambda=-1$. We identify two massive modes
$\Phi$ and $\Psi$ from the NMG (Fierz-Pauli action) using the
conventional black hole stability analysis in section 5. Finally,
we discuss similarity and difference in stability between the TMG
and the NMG in section 6.

We would like to mention the choice of a cosmological constant
$\Lambda$. In general, we choose $\Lambda=-1/\ell^2$ except
section 3 and 4, where  it is chosen to be $\Lambda=-1$ for the
simplicity of operator computation.

\section{ New massive gravity}

The NMG action~\cite{bht} composed of the
Einstein-Hilbert action with a cosmological constant $\lambda$ and
higher order curvature terms is given by
\begin{eqnarray}
\label{NMGAct}
 S^{(3)}_{NMG} &=& S^{(3)}_{EH}+S^{(3)}_{HC}, \\
\label{NMGAct2} S^{(3)}_{EH} &=& \frac{1}{16\pi G} \int d^3x \sqrt{-g}~ ( R-2\lambda),\\
\label{NMGAct3} S^{(3)}_{HC} &=& -\frac{1}{16\pi G m^2} \int d^3x
            \sqrt{-g}~\left(R_{\mu\nu}R^{\mu\nu}-\frac{3}{8}R^2\right),
\end{eqnarray}
where $G$ is a three-dimensional Newton's constant and $m^2$ a
parameter with mass dimension 2. From now on, we set $G = 1/8$ for
simplicity. The Einstein equation is given by \be \label{eineq}
G_{\mu\nu} +\lambda g_{\mu\nu}-\frac{1}{2m^2}K_{\mu\nu}=0, \ee where
the Einstein tensor $G_{\mu\nu}$  and $K_{\mu\nu}$ tensor are given
by
\begin{eqnarray}
  G_{\mu\nu}&=& R_{\mu\nu}-\frac{1}{2}g_{\mu\nu}R, \nonumber\\
  K_{\mu\nu}&=&2\nabla^2R_{\mu\nu}-\frac{1}{2}\nabla_\mu \nabla_\nu R-\frac{1}{2}\nabla^2Rg_{\mu\nu}\nonumber\\
        &+&4R_{\mu\rho\nu\sigma}R^{\rho\sigma} -\frac{3}{2} R R_{\mu\nu}-R_{\rho\sigma}R^{\rho\sigma}g_{\mu\nu}
         +\frac{3}{8}{R}^2 g_{\mu\nu}.
\end{eqnarray}
In order to have a black hole solution with dynamical exponent
$z$~\cite{z3,MyungD}, it is convenient to introduce dimensionless
parameters \be y=m^2~ \ell^2,~~w=\lambda~ \ell^2, \ee where $y$ and
$w$ are proposed to take \be
y=-\frac{z^2-3z+1}{2},~~w=-\frac{z^2+z+1}{2}. \ee For $z=1$
(nonrotating) BTZ black hole, one has $y=\frac{1}{2}$ and
$w=-\frac{3}{2}$, while $y=-\frac{1}{2}$ and $w=-\frac{13}{2}$ are
chosen for $z=3$ Lifshitz black hole.

In this work, we consider the BTZ black hole solution only
\begin{equation} \label{btz}
  ds^2_{\rm BTZ}=\bar{g}_{\mu\nu}dx^\mu dx^\nu=-\left(-{\cal M}+\frac{r^2}{\ell^2}\right)dt^2
   +\frac{dr^2}{\left(-{\cal M}+\frac{r^2}{\ell^2}\right)}+r^2d\phi^2,
\end{equation}
where ${\cal M}$ is the ADM mass determined to be ${\cal
M}=\frac{r_+^2}{\ell^2}$  with $r_+$ the horizon radius.
Importantly, the mass parameter $m^2$ and cosmological parameter
$\lambda$ are fixed as
\begin{equation}
m^2=\frac{1}{2\ell^2},~~\lambda=-\frac{3}{2\ell^2}
\end{equation}
to obtain the BTZ black hole.  In this background, taking into
account $\bar{g}^{\mu\nu}K_{\mu\nu}=2m^2 \lambda$,  the trace of
(\ref{eineq}) leads to the constant curvature scalar as
\begin{equation}
\bar{R}=4\lambda=-\frac{6}{\ell^2},
\end{equation} which is the same form as in the Einstein gravity
($R=6\Lambda$) with the cosmological constant $\Lambda=-1/\ell^2$. On
the other hand, the Ricci tensor takes the form
\begin{equation}
\bar{R}_{\mu\nu}=\lambda
\bar{g}_{\mu\nu}+\frac{1}{2m^2}K_{\mu\nu}=\frac{4}{3}\lambda
\bar{g}_{\mu\nu},
\end{equation}
which is the same as that of the Einstein gravity
\begin{equation}
\bar{R}_{\mu\nu}=2 \Lambda \bar{g}_{\mu\nu}. \end{equation} The
curvature tensor $\bar{R}_{\mu\rho\nu\sigma}$ takes the form
\begin{equation}
\bar{R}_{\mu\rho\nu\sigma}=\Lambda\Big(\bar{g}_{\mu\nu}\bar{g}_{\rho\sigma}-\bar{g}_{\mu\sigma}\bar{g}_{\rho\nu}\Big).\end{equation}
 For an AdS-sized black hole with $r_+=\ell$, one chooses the unit mass
of ${\cal M}=1$, which is designed for finding its quasinormal modes
(frequences). Setting $\ell^2=1~(\Lambda=-1)$, we rewrite the line
element (\ref{btz}) in global coordinates:
\begin{equation}
ds^2_{\rm M=1}=-\sinh^2(\rho)d\tau^2+\cosh^2(\rho)d\phi^2+d\rho^2,
\end{equation}
where the event horizon is located at $\rho=0~(r_+=1)$ while the
infinity is at $\rho=\infty~(r=\infty)$. The black hole temperature
is $T_H=1/4\pi$. The metric tensor $\bar{g}_{\mu\nu}$ can be  when
using the light cone coordinates $u/v=\tau\pm \phi$
\begin{equation} \bar{g}_{\mu\nu}=
\left(
  \begin{array}{ccc}
    \frac{1}{4} & -\frac{1}{4}\cosh(2\rho) & 0 \\
      -\frac{1}{4}\cosh(2\rho) & \frac{1}{4}  & 0 \\
    0 & 0 & 1 \\
  \end{array}
\right). \label{newm}
\end{equation}
Then the metric tensor (\ref{newm}) admits the Killing vector fields
$L_k$, $k=0,-1,1$ for local SL$(2,R)\times$SL$(2,R)$ algebra as
\begin{equation}
L_0=-\partial_u,~~L_{-1/1}=e^{\mp
u}\Big[-\frac{\cosh(2\rho)}{\sinh(2\rho)}\partial_u-\frac{1}{\sinh(2\rho)}\partial_v
\mp \frac{1}{2} \partial_\rho\Big],
\end{equation}
and $\bar{L}_0,\bar{L}_{-1/1}$ similarly by substituting
$u\leftrightarrow v$. Locally, they form a
basis of the SL$(2,R)$ Lie algebra  as
\begin{equation}
[L_0,L_{\pm 1}]=\mp L_{\pm 1},~~[L_1,L_{-1}]=2L_0, \end{equation}
which are useful for generating the whole tower of quasinormal
modes.

\section{Quasinormal modes}

Considering the perturbation $h_{\mu\nu}$ around the BTZ black hole
background $\bar{g}_{\mu\nu}$ in (\ref{newm})
\begin{equation}
g_{\mu\nu}=\bar{g}_{\mu\nu}+h_{\mu\nu}, \end{equation} the
linearized equation to (\ref{eineq}) takes the form \be
\label{keq}\delta G_{\mu\nu}(h)+\lambda h_{\mu\nu}-\frac{1}{2m^2}
\delta K_{\mu\nu}(h)=0, \ee where $\delta G_{\mu\nu}(h)$ and
$\delta K_{\mu\nu}(h)$ are the linearized Einstein tensor  and
linearized $K_{\mu\nu}$. We choose the transverse  and traceless
(TT) gauge to find a massive graviton propagation on the BTZ black
hole background as \be \label{g-f} \bar{\nabla}_\mu h^{\mu
\nu}=0,~~h\equiv h_\rho~^\rho=0. \ee We wish to mention that this
covariant gauge is also convenient for studying the conventional
stability of black holes in section 5. Then, the fourth order
equation (\ref{keq}) is split into \be \label{feq}
\Big[\bar{\nabla}^2-2\Lambda\Big]\Big[\bar{\nabla}^2-\Big(m^2+\frac{5\Lambda}{2}\Big)\Big]h_{\mu\nu}=0,\ee
which implies two branches of solutions~\cite{nmg-3}.  In this
section we choose $\Lambda=-1$ for simplicity of operator
calculation. The first equation is \be \label{seq1}
\Big[\bar{\nabla}^2-2\Lambda \Big]h^{L/R}_{\mu\nu}=0, \ee whose
solution  corresponds to the unphysical modes of  left- and
right-moving massless gravitons, while the second  is the equation
describing a physically massive graviton with 2 DOF \be
\label{seq2}
\Big[\bar{\nabla}^2-\Big(m^2+\frac{5\Lambda}{2}\Big)\Big]h^M_{\mu\nu}=0.
\ee In order to have a non-negative mass, one requires $m^2 \ge
1/2$. For $m^2=1/2$ (the BTZ black hole background), the fourth
order equation (\ref{feq}) could not be split into the two second
order equations (\ref{seq1}) and (\ref{seq2}).

First of all, we consider the AdS$_3$ background [or $M=-1$ in
(\ref{btz})]
\begin{equation} \label{ads3}
ds^2_{\rm AdS_3}=-\cosh^2(\rho)d\tau^2+\sinh^2(\rho)d\phi^2+d\rho^2,
\end{equation}
one can  easily  find the solution \cite{LSS} to (\ref{seq2}) as  \be
h^{M,AdS_3}_{\mu\nu}=e^{-ihu-i\bar{h}v}\frac{\sinh^2(\rho)}{(\cosh(\rho))^{h+\bar{h}}}
\left(
  \begin{array}{ccc}
    1 & \frac{h-\bar{h}}{2} & ia  \\
    \frac{h-\bar{h}}{2} & 1 & ib \\
    ia & ib & -a^2 \\
  \end{array}
\right), \label{adssol} \ee where $a$ and $b$ are given by \be
a=\frac{2}{\sinh(2\rho)},~~b=\frac{h-\bar{h}}{\sinh(2\rho)}.\ee We
have two sets for $h$ and $\bar{h}$ as primary states \be
h+\bar{h}=\frac{2+\sqrt{2+4m^2}}{2},~~h-\bar{h}=\pm 2, \ee where
$\pm$ denote  the solution to the first order equations
$(D^{M/\tilde{M}}h)_{\mu\nu}=0$ with the operators in Eq. (\ref{meq}),
respectively. This implies that the solutions to the first order
equations are also those to the second order equation.  In addition,
the solution to (\ref{seq1}) in the AdS$_3$ background (\ref{ads3})
has the same form as in (\ref{adssol}) when substituting
$(h,\bar{h})=(2,0)$ for left-moving mode and $(0,2)$ for
right-moving mode: these are also the solutions to
$(D^{L/R}h^{L/R})_{\mu\nu}=0$, respectively.

In this sense, hereafter, we use mainly the first order equations
instead of higher order equations to find quasinormal modes.  The
fourth order equation (\ref{feq}) can be expressed
\be \label{geq}
\Big(D^LD^RD^MD^{\tilde{M}}h\Big)_{\mu\nu}=0
\ee
in terms of mutually commuting operators as
\be \label{meq}
(D^{L/R})^\beta_\mu=\delta^\beta_\mu\pm
\epsilon_\mu~^{\alpha\beta}\bar{\nabla}_\alpha,~~(D^{M/\tilde{M}})^\beta_\mu=\delta^\beta_\mu\pm
\frac{1}{M} \epsilon_\mu~^{\alpha\beta}\bar{\nabla}_\alpha
\ee
with
\be \label{mass}
M=\sqrt{m^2+\frac{1}{2}}.
\ee
At the critical point
(the BTZ black hole background: $m^2=1/2,~M=1)$, the operators $D^M$ and
$D^L$ degenerate, while the operators $D^{\tilde{M}}$ and $D^R$
degenerate. Thus, we have to treat it separately in the next
section.

On the other hand, the second order massive equation (\ref{seq2})
can be expressed as~\cite{LS}
\be \label{2ndm}
(D^MD^{\tilde{M}}h)_{\mu\nu}=0.
\ee
In order to find
quasinormal modes, we have to solve (\ref{2ndm}) together with the
TT gauge (\ref{g-f}). This approach was used to derive quasinormal
modes in the TMG~\cite{Sachs}, which shows the quasinormal modes of a
minimally coupled scalar.  We remind the reader that the solution to
(\ref{2ndm}) may be a linear combination of two solutions to the
first order equations \ba
\label{eql1} (D^Mh)_{\mu\nu}&=&0\to \epsilon_\mu~^{\alpha\beta}\bar{\nabla}_\alpha h_{\beta\nu}+Mh_{\mu\nu}=0,\\
\label{eql2} (D^{\tilde{M}}h)_{\mu\nu}&=&0 \to
\epsilon_\mu~^{\alpha\beta}\bar{\nabla}_\alpha
h_{\beta\nu}-Mh_{\mu\nu}=0 \ea together with the TT gauge
(\ref{g-f}). Let us first find the right-moving solution to
(\ref{eql1}). The least damped ($n=0$) quasinormal mode to
(\ref{eql1}) could be found by considering  the form \be
\label{hmassive}
h^M_{\mu\nu}=e^{-i(\omega\tau+k\phi)}\psi^M_{\mu\nu}(\rho)=e^{-ip_+u-ip_-v}\psi^M_{\mu\nu}(\rho),~~p_+\pm
p_-=\omega/k,\ee where \be \psi^M_{\mu\nu}(\rho)=F^M(\rho)\left(
  \begin{array}{ccc}
    1 & 0 & \frac{2}{\sinh(2\rho)}\\
    0 & 0 & 0 \\
    \frac{2}{\sinh(2\rho)} & 0 & \frac{4}{\sinh^2(2\rho)}\\
  \end{array}
\right). \label{mhme} \ee The transversality condition of
$\bar{\nabla}_\mu h^{\mu\nu}=0$ implies the chiral highest weight
condition of $\bar{L}_{1}h_{\mu\nu}=0$ under the form of
$h^M_{\mu\nu}$ in (\ref{hmassive}), giving the constraint  \be
\Big[2i
p_++2ip_-\cosh(2\rho)+\sinh(2\rho)\partial_\rho\Big]F^M(\rho)=0,
\ee whose solution is given by \be
F^M(\rho)=[\sinh(\rho)]^{-2ip_-}[\tanh(\rho)]^{-ik}. \ee From
(\ref{eql1}) for $\mu=\nu=\rho$,  $p_-$ is determined to be \be
p_-=-i h_R(M),~~h_R(M)=-\frac{M}{2}-\frac{1}{2}. \ee At the first
sight, the $n=0$ quasinormal mode seems to be  \be \label{psim}
h^M_{\mu\nu}=e^{-ik(\tau+\phi)-2h_R(M)\tau} \psi^M_{\mu\nu}(\rho).
\ee Considering the form of quasinormal frequency \be
\omega=\omega_R-i\omega_I,~~ \omega_I>0\ee we read off it from
(\ref{psim}) \be \omega_M=k-2i h_R(M). \ee We note here that
``$k$" is the angular quantum number as well as the real part of
quasinormal frequency.  In order that $\omega_M$ be a quasinormal
mode, $h_R(M)$ is required to be positive and thus, one has to
have $M<-1$. However, it is impossible to make $M<-1$ because
$M\ge 1$, as is shown in (\ref{mass}).  Hence it is clear that
(\ref{psim}) is not considered as a right-moving quasinormal mode
for the least damped case.

The next step is to find the right-moving solution to (\ref{eql2}).
This solution is obtained simply  by replacing $M$ by $\tilde{M}=-M$
[equivalently, $h_R(M)$ by $h_R(\tilde{M})$] in (\ref{psim}) as \be
\label{psim2}
h^{\tilde{M}}_{\mu\nu}=e^{-ik(\tau+\phi)-2h_R(\tilde{M})\tau}
\psi^{\tilde{M}}_{\mu\nu}(\rho), \ee where \be
h_R(\tilde{M})=\frac{M}{2}-\frac{1}{2},~~M\ge 1. \ee It seems  that
(\ref{psim2})  is regarded as
 a right-moving quasinormal mode of the least damped case whose quasinormal frequency is given by
\be \omega_{\tilde{M}}=k-2ih_R(\tilde{M}). \ee Hence, we could
construct the overtone quasinormal modes for this solution.
Introducing the operator combination $L_{-1}\bar{L}_{-1}$ which
replaces $\omega_I$ by $\omega_I-2$, the $n^{\rm th}$-overtone
quasinormal mode is constructed by \be \label{psima}
h^{\tilde{M},n}_{\mu\nu}=\left(L_{-1}\bar{L}_{-1}\right)^n
h^{\tilde{M}}_{\mu\nu} \ee whose quasinormal frequency takes the
form \be
\label{qnormall}\omega_{\tilde{M}}^n=k-2i\Big[h_R(\tilde{M})+n\Big],~~
n \in Z.\ee

Similarly,
 when imposing  the anti-chiral highest weight condition of $L_1h^M_{\mu\nu}=0$ and thus, $p_+=-ih_L(M)$,
  the left-moving  quasinormal modes to (\ref{eql1}) take the form \be \label{psimb}
h^{M,n}_{\mu\nu}=
(L_{-1}\bar{L}_{-1})^nh^{M}_{\mu\nu}=e^{ik(\tau-\phi)-2h_L(M)\tau}
(L_{-1}\bar{L}_{-1})^n\psi^{M}_{\mu\nu}(\rho), \ee where \be
\psi^{M}_{\mu\nu}(\rho)=[\sinh(\rho)]^{-2h_L(M)}[\tanh(\rho)]^{ik}\left(
  \begin{array}{ccc}
   0 & 0 & 0 \\
    0 & 1 & \frac{2}{\sinh(2\rho)}\\
    0& \frac{2}{\sinh(2\rho)}  & \frac{4}{\sinh^2(2\rho)}\\
  \end{array}
\right),~~h_L(M)=\frac{M}{2}-\frac{1}{2}, \label{mhmeb}\ee whose
quasinormal modes are given by
 \be \label{qnormalr}
\omega_{M}^n=-k-2i\Big[h_L(M)+n\Big],~~ n \in Z.\ee We note that two
quasinormal modes $h^{\tilde{M}}_{\mu\nu}$ and $h^{M}_{\mu\nu}$ are
ingoing (the right- and left-moving)  modes at the horizon $\rho=0$ and
are normalizable modes at $\rho=\infty$ satisfying the Dirichlet
boundary condition. It is worth noting that
 the left-moving quasinormal modes to
(\ref{eql2}) with \be h_L(\tilde{M})=-\frac{M}{2}-\frac{1}{2}\ee is
not available since $h_L(\tilde{M})<0$ due to $M\ge 1$. We summarize
the possible quasinormal modes in the following Table.
\begin{table}[ht]
\label{table} \centering
\begin{tabular}{ccc}
\hline\hline
equation & right-moving (R) mode & left-moving (L) mode \\
\hline\hline
(\ref{eql1}) & $h^{M,R}_{\mu\nu}$~(X) & $h^{M,L}_{\mu\nu}$~(O) \\
 (\ref{eql2}) & $h^{\tilde{M},R}_{\mu\nu}$~(O) & $h^{\tilde{M},L}_{\mu\nu}$~(X) \\
\hline
\end{tabular}
\end{table}
In the BTZ black hole background, (\ref{eql1}) allows the L-mode
only, while (\ref{eql2}) does the R-mode.

In this section, we have obtained the L/R-moving quasinormal
modes, which show at least the $s$-mode stability using
(\ref{unstablec}): for $k=0~(\omega_R=0)$, one has
$\omega^n_{I,\tilde{M}/M}=2n+h_{R/L}(\tilde{M}/M)>0$. In this
case, $k=0$ requires the disappearance of the L/R-moving modes.

At this stage, it seems appropriate to comment
on the similarity and difference between
TMG and NMG. In deriving the quasinormal modes, we have used the
operator method. As far as this method is concerned, there is no
essential difference between TMG and NMG. The difference is that
the substitution of $M \to \mu$ in (\ref{eql1}) is necessary for
the TMG and (\ref{eql2}) is required additionally for the NMG
because the parity is preserved in the NMG. However, we note that
the L/R-moving modes (\ref{psima}) and (\ref{psimb}) are not
orthogonal to each other even $s$-mode ($k=0$) in the NMG because
they contain $h_{\rho\rho}$ commonly. This gives rise to a
significant difference in obtaining the quasinormal modes of the
same BTZ black hole between two massive gravity theories.

Finally,  at the chiral point of $M=1$, two $h^{\tilde{M}}_{\mu\nu}$
and $h^{M}_{\mu\nu}$ are no longer $n=0$ quasinormal modes because
of $\omega_I=0~(h_{L/R}(1)=0)$. Hence, we will treat it in the next
section. In addition, we mention that requiring both of
$\bar{L}_{-1} h_{\mu\nu}=0$ and $L_{-1} h_{\mu\nu}=0$ lead to normal
modes with ingoing and outgoing fluxes at the event horizon as in
Ref.~\cite{Birm}, which is not the condition for obtaining
quasinormal modes.

\section{Quasinormal modes at the critical point}
In order to see what happens in the perturbation at the critical
point (the BTZ black hole background), we  need to solve  the
fourth order linearized equation.  In this section we choose
$\Lambda=-1$ for simplicity of operator computation. In general, a
mode annihilated by $D^M(D^L)[D^R]\{(D^L)^2$ but not by $D^L\}$ is
called massive (left-moving) [right-moving] \{left-logarithmic
(L,new)\} and is denoted by
$h^M_{\mu\nu}(h^L_{\mu\nu})[h^R_{\mu\nu}]\{h^{L,new}_{\mu\nu}\}$~\cite{Grum}.
Also, a mode annihilated by $(D^R)^2$ but not by $D^R$ is called
right-logarithmic mode and is denoted by $h^{R,new}_{\mu\nu}$.
Away from the critical point ($m^2\not=1/2$), the general solution
to (\ref{geq}) is obtained by linearly combining left, right, and
two massive modes.

At the critical point, $m^2=1/2$, $D^M$ degenerates into $D^L$, while
$D^{\tilde{M}}$ degenerates into $D^R$. The L/R-moving modes are
purely gauge degrees of freedom (unphysical modes), whereas  two
massive modes and their logarithmic partners  constitute physically
propagating bulk modes. At the critical point (the BTZ black hole
background), the fourth order equation (\ref{feq}) becomes~\cite{GH}
\be \label{ceq} (D^RD^L)^2h^{L/R,new}_{\mu\nu}=0 \to
\Big[\bar{\nabla}^2-2\Lambda\Big]^2h^{L/R,new}_{\mu\nu}=0, \ee which
is basically different from the second order equations (\ref{seq1})
and (\ref{seq2}).   The reason is clear when observing
\be\label{2ndeqs} (D^RD^Lh^{L/R,new})_{\mu\nu}=-2h^{L/R}_{\mu\nu}
\to
\Big[\bar{\nabla}^2-2\Lambda\Big]h^{L/R,new}_{\mu\nu}=2h^{L/R}_{\mu\nu},
\ee which was derived in Appendix I in detail. See Ref.~\cite{GH} for
the derivation this relation  on AdS$_3$ background.  This naturally
leads to (\ref{ceq}) when operating $(\bar{\nabla}^2-2\Lambda)$ on
both sides and using (\ref{seq1}).
 In this case, considering $M\to L$ and $\tilde{M}\to R$, one has
 \begin{eqnarray}
 h^{L}_{\mu\nu} &=& e^{ik(\tau -\phi)}\psi^{L}_{\mu\nu}(\rho) \nonumber\\
 &=& e^{ik(\tau -\phi)} [\tanh(\rho)]^{ik}
  \left(
  \begin{array}{ccc}
   0 & 0 & 0 \\
    0 & 1 & \frac{2}{\sinh(2\rho)}\\
    0& \frac{2}{\sinh(2\rho)}  & \frac{4}{\sinh^2(2\rho)}\\
  \end{array}
 \right),  \\
 h^{R}_{\mu\nu}&=& e^{-ik(\tau+\phi)}\psi^{R}_{\mu\nu}(\rho)
 \nonumber\\
\label{logeom2}  &=& e^{-ik(\tau+\phi)} [\tanh(\rho)]^{-ik} \left(
  \begin{array}{ccc}
    1 & 0 & \frac{2}{\sinh(2\rho)}\\
    0 & 0 & 0 \\
    \frac{2}{\sinh(2\rho)} & 0 & \frac{4}{\sinh^2(2\rho)}\\
  \end{array}
  \right).
\end{eqnarray}
Then, we construct two newly logarithmic solutions of L/R modes to
(\ref{ceq}) \ba\label{logeom1}
h^{L,new}_{\mu\nu}&=&\partial_{m^2}h^{M}_{\mu\nu}|_{m^2=1/2}=y(\tau,\rho)h^{L}_{\mu\nu},\\
h^{R,new}_{\mu\nu}&=&\partial_{m^2}h^{\tilde{M}}_{\mu\nu}|_{m^2=1/2}=y(\tau,\rho)h^{R}_{\mu\nu},
\ea where $y(\tau,\rho)$ is defined by \be y(\tau,\rho)=-\tau
-\ln[\sinh(\rho)]. \ee

We are now in a position to  construct two quasinormal modes in the
BTZ black hole background  inspired by left-logarithmic sector of
TMG. Unfortunately, $h^{L,new}_{\mu\nu}$  is growing in time $\tau$
and $\rho$, showing disqualification as a quasinormal mode. Hence,
Sachs~\cite{Sachs2} has  proposed that the left-logarithmic
quasinormal mode can be constructed by considering
\be\label{higherDSs}
h^{L,(n)}_{\mu\nu}=\Big[\bar{L}_{-1}L_{-1}\Big]^n
h^{L,new}_{\mu\nu}. \ee  For example, a candidate for quasinormal
mode  is the first descendent given by
\begin{eqnarray}
\label{1stL}
 h^{L,(1)}_{\mu\nu} &=& (\bar{L}_{-1}L_{-1}) h^{L,new}_{\mu\nu} \nonumber\\
 &=& \left[\frac{1}{2}-y(\tau,\rho)+\frac{ik}{2}
                  y(\tau,\rho)\right]\psi^{L,(1)}_{\mu\nu} \nonumber\\
                 &+&  \frac{2e^{-2\tau}}{\sinh^2(\rho)} \left[\cosh^2(\rho)+\frac{1}{2}k^2 y(\tau,\rho)
                  + ik\left(y(\tau,\rho)-\frac{1+\cosh^2(\rho)}{2}\right)\right]h^{L}_{\mu\nu},
\end{eqnarray}
as the left-logarithmic solution. Here \be
\psi^{L,(1)}_{\mu\nu}=\frac{2e^{-2\tau}}{\sinh^2(\rho)}e^{ik(\tau-\phi)}(\tanh(\rho))^{ik}
\left(\begin{array}{ccc}
             0 & 1 & \frac{2}{\sinh(2\rho)} \\
             1 & 1 & \frac{2\cosh(\rho)}{\sinh(\rho)} \\
             \frac{2}{\sinh(2\rho)} & \frac{2\cosh(\rho)}{\sinh(\rho)} & \frac{4(1+2\cosh(2\rho))}{\sinh^2(2\rho)} \\
\end{array}
\right). \ee where we observe that $h^{L,(1)}_{\mu\nu}$ shows a
genuine quasinormal mode with exponential fall-off in $\tau$ and
$\rho$. Note that in the limit of $k\to 0$,  the first descendent
becomes (2.18) of Ref.~\cite{Sachs2} for the TMG without
$\psi^{L}_{\mu\nu}$ which may be considered as as pure
gauge~\footnote{We thank I. Sachs for pointing out it.}. Here we
emphasize again that the $k\to 0$ limit denotes both the $s$-mode
of $\phi$ and $\omega_R=0$, which confirms the $s$-left
logarithmic mode stability of BTZ black hole. The higher
descendent modes in principle can be obtained by using
(\ref{higherDSs}), however, it seems to be a formidable  task.
Thus, we confine ourselves  to the $k=0$ ($s$-mode) case, for
simplicity, and obtain the second descendent mode as
\begin{equation}
\label{2ndL}
 h^{L,(2)}_{\mu\nu}=\Big[\bar{L}_{-1}L_{-1}\Big]^2 h^{L,new}_{\mu\nu}
      =\frac{e^{-4\tau}}{\sinh^4\rho}\Bigg[\psi^{L,(2A)}_{\mu\nu}-12y(\tau,\rho)\psi^{L,(2B)}_{\mu\nu}\Bigg],
\end{equation}
where
\begin{eqnarray}
  \psi^{L,(2A)}_{\mu\nu} &=&
  \left(\begin{array}{ccc}
      6 & 11+9\cosh(2\rho) & \frac{2(11+15\cosh(2\rho))}{\sinh(2\rho)} \\
     11+9\cosh(2\rho) & 11+9\cosh(2\rho) & 4(7+3\cosh(2\rho))\coth(\rho) \\
      \frac{2(11+15\cosh(2\rho))}{\sinh(2\rho)}  &  4(7+3\cosh(2\rho))\coth(\rho) & \frac{4(26+31\cosh(2\rho)+9\cosh(4\rho))}{\sinh^2(2\rho)} \\
  \end{array}\right),\nonumber\\
    \psi^{L,(2B)}_{\mu\nu} &=&
  \left(\begin{array}{ccc}
      1 &  2+\cosh(2\rho) & 4\coth(\rho) \\
     2+\cosh(2\rho) & 2+\cosh(2\rho) & (5+\cosh(2\rho))\coth(\rho) \\
      4\coth(\rho) & (5+\cosh(2\rho))\coth(\rho)  & \frac{2(3+2\cosh(2\rho))}{\sinh^2(\rho)} \\
  \end{array}\right).
\end{eqnarray}
Restoring the $k$-dependence, we may  find by induction that the
higher modes $h^{R,(n)}_{\mu\nu} ~(n\ge 2)$ are constructed by
replacing $e^{-2\tau}$ by $e^{-2n\tau}$.  Then, the quasinormal
frequencies may be  given by \be \omega^n_L=-k-2i n, ~~n \in Z, \ee
which are the same quasinormal frequencies of left-moving modes
$\omega^n_{\tilde{M}}$ with $h_R(\tilde{M}=1)=0$ in
(\ref{qnormall}).

The right-logarithmic  quasinormal modes could be  similarly
constructed by
\begin{eqnarray}
\label{1stR}
 h^{R,(1)}_{\mu\nu} &=& (\bar{L}_{-1}L_{-1}) h^{R,new}_{\mu\nu} \nonumber\\
                  &=&\left[\frac{1}{2}-y(\tau,\rho)-\frac{i k}{2}
                                      y(\tau,\rho)\right] \psi^{R,(1)}_{\mu\nu}   \nonumber\\
                  &+& \frac{2e^{-2\tau}}{\sinh^2(\rho)}\left[\cosh^2(\rho)+\frac{1}{2}k^2 y(\tau,\rho)
                      - ik\left(y(\tau,\rho)-\frac{1+\cosh^2(\rho)}{2}\right)\right]h^{R}_{\mu\nu},
\end{eqnarray}
where
 \be \psi^{R,(1)}_{\mu\nu}
 =\frac{2e^{-2\tau}}{\sinh^2(\rho)}e^{-ik(\tau+\phi)}(\tanh(\rho))^{-ik}
\left(\begin{array}{ccc}
 1 & 1 & \frac{2\cosh(\rho)}{\sinh(\rho)}  \\
 1 & 0 & \frac{2}{\sinh(2\rho)} \\
 \frac{2\cosh(\rho)}{\sinh(\rho)} & \frac{2}{\sinh(2\rho)} & \frac{4(1+2\cosh(2\rho))}{\sinh^2(2\rho)} \\
\end{array}
   \right).\ee
Here we observe that $h^{R,(1)}_{\mu\nu}$  is a genuine quasinormal
mode with exponential fall-off in $\tau$ and $\rho$. Considering the
$k =0$ ($s$-mode) case, for simplicity, the second descendent mode
is given by
\begin{equation}
\label{2ndR}
 h^{R,(2)}_{\mu\nu}=\Big[\bar{L}_{-1}L_{-1}\Big]^2 h^{R,new}_{\mu\nu}
      =\frac{e^{-4\tau}}{\sinh^4\rho}\Bigg[2\psi^{R,(2A)}_{\mu\nu}-12y(\tau,\rho)\psi^{R,(2B)}_{\mu\nu}\Bigg],
\end{equation}
where
\begin{eqnarray}
  \psi^{R,(2A)}_{\mu\nu} &=&
  \left(\begin{array}{ccc}
      10+7\cosh(2\rho)  & 9+3\cosh(\rho)+5\cosh(2\rho) & \frac{31+33\cosh(2\rho)+4\cosh(4\rho)}{\sinh(2\rho)} \\
    9+3\cosh(\rho)+5\cosh(2\rho) & 6 & \frac{21+25\cosh(2\rho)}{\sinh(2\rho)} \\
      \frac{31+33\cosh(2\rho)+4\cosh(4\rho)}{\sinh(2\rho)}   & \frac{21+25\cosh(2\rho)}{\sinh(2\rho)}  & \frac{2P(\rho)}{\sinh^2(2\rho)} \\
  \end{array}\right),\nonumber\\
    \psi^{R,(2B)}_{\mu\nu} &=&
  \left(\begin{array}{ccc}
      2+\cosh(2\rho)  &  2+\cosh(2\rho) & (5+\cosh(2\rho))\coth(\rho) \\
     2+\cosh(2\rho) & 1 & 4\coth(\rho) \\
      (5+\cosh(2\rho))\coth(\rho) & 4\coth(\rho)  & \frac{2(3+2\cosh(2\rho))}{\sinh^2(\rho)} \\
  \end{array}\right)
\end{eqnarray}
with
\begin{eqnarray}
 P(\rho)=36+4\cosh(\rho)+40\cosh(2\rho)+10\cosh(4\rho)+6\sinh(2\rho))+3\sinh(4\rho).
\end{eqnarray}
Assuming the $k$-dependence again,  the higher modes
$h^{R,(n)}_{\mu\nu}~(n\ge 2)$ can be obtained by replacing
$e^{-2\tau}$ by $e^{-2n\tau}$ as well. Thus, the quasinormal
frequencies may be  given by \be \omega^n_L=k-2i n, ~~n \in Z, \ee
which are the same quasinormal frequencies of right-moving modes
$\omega^n_M$ with $h_L(M=1)=0$ in (\ref{qnormalr}).

 In this section,
we have obtained L/R-logarithmic quasinormal modes, which show at
least the $s$-mode stability using (\ref{unstablec}): for
$k=0~(\omega_R=0)$, one has $\omega^n_{I,L/R}=2n>0$.

Finally, we note the similarity  and difference between TMG and
NMG. In deriving the quasinormal modes at the critical point, we
have used the operator method. There is no essential difference
between TMG and NMG. The difference is that  the L-logarithmic
mode approach is necessary for the TMG, while both of L/R
logarithmic approaches are  required  for the NMG because the
parity is preserved in the NMG.  However, it is not clear  that
two L/R-logarithmic quasinormal modes are  orthogonal to each
other even for the $s$-mode ($k=0$).  This is an  important issue
because the orthogonality may guarantee two independent massive
quasinormal modes at the critical point.

\section{Two $s$-massive modes in NMG}
First of all, we point out a few problems on the previous tensor
quasinormal modes. Their frequencies were constructed as those of a
minimally coupled scalar mode. The L/R modes of (\ref{psima}) and
(\ref{psimb}) are not orthogonal to each other even for $s$-mode
($k=1$) because they contain $h_{\rho\rho}$ commonly. Hence, we did
not confirm that the L/R-moving modes are two truly propagating
modes in the new massive gravity. This asks directly what is the
explicit form of two physically massive modes propagating in the BTZ
black hole. In this section, we wish to identify two massive
graviton modes propagating in the BTZ black hole background.

In order to perform the stability analysis of massive graviton with
2 DOF, we first recall Eq. (\ref{btz}) as the BTZ black hole
background.  In this background the perturbation equations
(\ref{seq2}) can be rewritten as
\begin{eqnarray}\label{btz1}
\bar{\nabla}^2_{\rm BTZ}
h_{\mu\nu}-\Bigg[m^2+\frac{5\Lambda}{2}\Bigg]h_{\mu\nu}=0.
\end{eqnarray}
This equation is similar to the Fierz-Pauli massive equation
\begin{eqnarray}\label{btzFP}
\bar{\nabla}^2_{\rm BTZ}
h_{\mu\nu}-\Bigg[m^2+2\Lambda\Bigg]h_{\mu\nu}=0,
\end{eqnarray}
which is  derived from Fierz-Pauli action together with the TT
gauge. In order to solve (\ref{btz1}) explicitly with the TT gauge,
we consider the following two distinct perturbing metric ansatz: the
 type I  has  two off-diagonal components $h_0$ and $h_1$
\begin{eqnarray}
h^I_{\mu\nu}=\left(
\begin{array}{ccc}
0 & 0 & h_0(r) \cr 0 & 0 & h_1(r) \cr h_0(r) & h_1(r) & 0
\end{array}
\right) e^{\omega_h t}e^{ik\phi} \,, \label{oddp}
\end{eqnarray}
while for the type II, the metric tensor takes the form with four
components $H_0,~H_1,~H_2,$ and $H_3$ as~\cite{LW}
\begin{eqnarray}
h^{II}_{\mu\nu}=\left(
\begin{array}{ccc}
H_0(r)  & H_1(r) & 0  \cr H_1(r) & H_2(r) & 0 \cr 0 & 0 & H_3(r)
\end{array}
\right) e^{\omega_h t}e^{ik\phi} \,. \label{evenp}
\end{eqnarray}
At this stage, we note that the above two choices of (\ref{oddp})
and (\ref{evenp}) are working only for $s$-mode ($k=0$)
perturbation. Also, two are orthogonal to each other.  In  Appendix
II, we have shown that $k=0$ case leads to two consistent decoupling
processes for I and II metric perturbations, when solving
(\ref{btz1}).
\begin{figure*}[t!]
   \centering
   \includegraphics{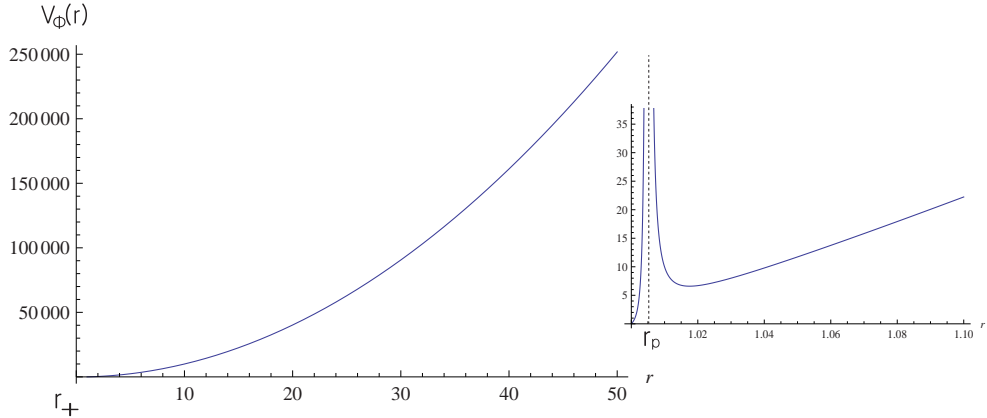}
\caption{$V_\Phi$ graphs as function of $r$ with fixed values
$\ell^2=1,~{\cal M}=1,~\omega=1$, and $m=10$. For $m^2\neq1/2\ell^2$
case, $V_\Phi$ blows up at $r=r_p=\sqrt{l^2{{\cal
M}+\omega^2/(m^2-1/2\ell^2)}}(\approx 1.005$, near the event
horizon), while for the critical $m^2=1/2\ell^2$ case, there is no
the blowing-up point.}
\end{figure*}

Substituting Eq. (\ref{oddp}) into Eq. (\ref{btz1}) and eliminating
$h_1$ from $(t,\phi$) and  $(r,\phi)$ components of (\ref{btz1}), we
obtain
\begin{eqnarray}\label{oddeq}
&&\hspace{-5em}\left\{(m^2-1/2\ell^2)(r^2/\ell^2-{\cal
M})+\omega_h^2\right\}h_0'' +\left\{\frac{r^2/\ell^2+{\cal
M}}{r(r^2/\ell^2-{\cal M})}\omega_h^2-\frac{r^2/\ell^2-{\cal
M}}{r}(m^2-1/2\ell^2)\right\}h_0'
\nonumber\\
&&\hspace{3em}-\left\{\frac{4\omega_h^2}{\ell^2(r^2/\ell^2-{\cal
M})}+\left(m^2-1/2\ell^2 +\frac{\omega_h^2}{r^2/\ell^2-{\cal
M}}\right)^2\right\}h_0=0.
\end{eqnarray}
Note that we consider only $s$-mode ($k=0$) for components
($t,r$),($r,r$) of Eq. (\ref{btz1}). At the horizon $r=r_+$ and the
spatial infinity $r=\infty$, a solution to the above equation
behaves as
\begin{eqnarray}
  h_0\sim
(r-r_+)^{\pm\frac{\omega_h}{2\sqrt{{\cal M}/\ell^2}}}=e^{\pm
\frac{\omega_h}{2\sqrt{{\cal M}/\ell^2}}\ln[r-r_+]},~~ {\rm
and}~~h_0\sim r^{1\pm\sqrt{\ell^2m^2+1/2}},
\end{eqnarray}
where $r_+=\sqrt{{\cal M}\ell^2}$. Introducing the tortoise
coordinate $r^*$ in [$dr^*=dr/(-{\cal M}+r^2/\ell^2)$], and
redefining $\omega_h=-i\omega$ and a new field $\Phi$ as
\begin{equation}
\Phi=\frac{h_0}{g(r)} \end{equation}
 with
\begin{equation}
g(r)=\sqrt{r\left\{(m^2-1/2\ell^2)r^2/\ell^2-(m^2-1/2\ell^2){\cal
M}+\omega_h^2\right\}}~, \end{equation} Eq.(\ref{oddeq}) can be
written as the Schr\"odinger-type equation
\begin{eqnarray} \label{schphi}
\frac{d^2\Phi}{dr^{*2}}+[\omega^2-V_\Phi(\omega,r)]\Phi=0,
\end{eqnarray}
where the $\omega$-dependent potential takes the form
\begin{eqnarray}
&&V_\Phi(\omega,r)=(r^2/\ell^2-{\cal
M})\Bigg[m^2+\frac{13}{4\ell^2}-\frac{3{\cal M}}{4r^2}+
\frac{3(m^2-1/2\ell^2)^2 r^2(r^2/\ell^2-{\cal
M})}{\ell^4\left\{(m^2-1/2\ell^2)(-{\cal
M}+r^2/\ell^2)-\omega^2\right\}^2}\nonumber\\
&&\hspace{11em} +\frac{2(m^2-1/2\ell^2)(2{\cal
M}-3r^2/\ell^2)}{\ell^2\left\{(m^2-1/2\ell^2)(-{\cal
M}+r^2/\ell^2)-\omega^2\right\}}\Bigg].
\end{eqnarray}
It is important to note that for $m^2\geq1/2\ell^2$, the above
potential is always positive for whole range of $r_+\leq r^*\leq
\infty$  even though it blows up at $r=r_p$ (see Fig.1), which
implies that the type I-perturbation is stable.

On the other hand, plugging Eq.(\ref{evenp}) into Eq.(\ref{btz1})
and after manipulations we obtain the second differential equation
for $H_1(r)$ as
\begin{eqnarray}\label{eveneq}
&&\left\{(m^2-1/2\ell^2)(r^2/\ell^2-{\cal M})+r^2/\ell^4-2{\cal
M}/\ell^2+\omega_h^2\right\}H_1''+\left\{ \frac{7r^2/\ell^2-{\cal
M}}{r(r^2/\ell^2-{\cal
M})}\omega_h^2\right.\nonumber\\
&&\left.+\frac{5r^2/\ell^2-{\cal M}}{r}(m^2-1/2\ell^2)
+\frac{5r^4/\ell^4-13{\cal M}r^2/\ell^2+2{\cal M}^2}{\ell^2
r(r^2/\ell^2-{\cal M})}\right\}H_1'+ \left\{\frac{6r^4/\ell^4-{\cal
M}^2}{r^2(r^2/\ell^2-{\cal M})^2}\omega_h^2
\right.\nonumber\\
&&\left.+\frac{2r^4/\ell^4-2{\cal M}r^2/\ell^2-{\cal
M}^2}{r^2(r^2/\ell^2-{\cal M})}(m^2-1/2\ell^2)
+\frac{(3r^2/\ell^2-2{\cal M})(r^4/\ell^4-4{\cal M}r^2/\ell^2-{\cal
M}^2)}{\ell^2r^2(r^2/\ell^2-{\cal M})^2}
\right.\nonumber\\
&&\left.\hspace{3em}-\left(m^2-1/2\ell^2+\frac{\omega_h^2}{r^2/\ell^2-{\cal
M}}\right)^2 \right\}H_1=0.
\end{eqnarray}
In this case, we also consider only $s$-mode ($k=0$) for component
($t,\phi$) of Eq.(\ref{btz1}). We have wave forms near the horizon
and spatial infinity
\begin{eqnarray}
H_1\sim (r-r_+)^{-1\pm\frac{\omega_h}{2\sqrt{{\cal M}/\ell^2}}}=e^{[
-1\pm \frac{\omega_h}{2\sqrt{{\cal M}/\ell^2}}]\ln[r-r_+]},~~{\rm
and}~~H_1\sim r^{-2\pm\sqrt{\ell^2 m^2+1/2}}.
\end{eqnarray}
\begin{figure*}[t!]
   \centering
   \includegraphics{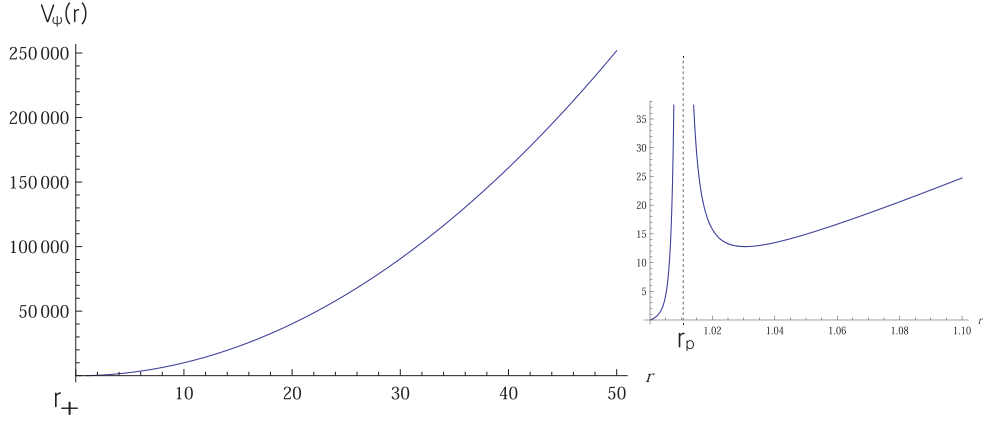}
\caption{$V_\Psi$ graphs as function of $r$ with fixed values
$\ell^2=1,~{\cal M}=1,~\omega=1$, and $m=10$. In right graph,
$V_\Psi$ blows up at $r=r_p=\sqrt{\ell^2{{\cal M}+({\cal
M}/\ell^2+\omega^2)/(m^2+1/2\ell^2)}}(\approx 1.01$ near the even
horizon $r=r_+$).}
\end{figure*}
Introducing the tortoise coordinate $r^*$, and redefining
$\omega_h=-i\omega$ and  a new field $\Psi$
 \begin{equation}
 \Psi=\frac{H_1}{f(r)}\end{equation}
  with
\begin{equation}
f(r)=\frac{\sqrt{{\cal M}(m^2-1/2\ell^2)+2{\cal
M}/\ell^2-(m^2-1/2\ell^2)r^2/\ell^2-r^2/\ell^4-\omega_h^2}}
{\sqrt{r(r^2/\ell^2-{\cal M})^2}},\end{equation}  Eq.(\ref{eveneq})
can be written as the Schr\"odinger-type equation
\begin{eqnarray} \label{schpsi}
\frac{d^2\Psi}{dr^{*2}}+[\omega^2-V_\Psi(\omega,r)]\Psi=0,
\end{eqnarray}
where the $\omega$-dependent potential is given by
\begin{eqnarray}
&&\hspace{-2em}V_\Psi(\omega,r)=(r^2/\ell^2-{\cal
M})\Bigg[m^2+\frac{5}{4\ell^2}-\frac{3{\cal M}}{4r^2}
 +\frac{3r^2(m^2+1/2\ell^2)^2(r^2/\ell^2-{\cal
M})} {\ell^4\left({\cal M}m^2+3{\cal
M}/2\ell^2-m^2r^2/\ell^2-r^2/2\ell^4+\omega^2\right)^2}
\nonumber\\&&\hspace{8em}+\frac{4(m^2+1/2\ell^2)(r^2/\ell^2-{\cal
M})} {\ell^2\left ({\cal M}m^2+3{\cal
M}/2\ell^2-m^2r^2/\ell^2-r^2/2\ell^4+\omega^2\right)}\Bigg].
\end{eqnarray}
It is important to note that for $m^2\geq1/2\ell^2$, the above
potential is always positive for whole range of $r_+\le r\le
\infty~(-\infty\leq r^*\leq 0)$ (see Fig.2) even though it blows up
at $r=r_p$, which implies that the type II-perturbation is stable.
This has been  confirmed by computing quasinormal modes numerically
when using (\ref{eveneq}) in Ref.~\cite{LW}, instead of (\ref{schpsi}).

Finally, we have shown that the positivity of two potentials leads
to the stability condition of the BTZ black hole
\begin{equation} \label{stcon}
m^2 \ge \frac{1}{2\ell^2},
\end{equation}
which is consistent with the condition of $M \ge \ell$.

\section{Discussions}
We have performed the stability analysis of the BTZ black hole in
the NMG.  We have derived quasinormal modes of the NMG by solving the
first order equations, which is based on Sachs and Solodukhin
method used in the TMG.  However, we have observed that the  left- and
right-modes are not orthogonal, which has a problem to be considered
as two independent massive modes.  Hence, we have identified two
massive modes from the NMG (Fierz-Pauli action) using the conventional
black hole stability analysis. Furthermore, we have obtained left-
and right logarithmic  quasinormal modes at the critical point of
$m^2\ell^2=1/2$ using the LCFT.

We discuss similarity and difference in stability between the TMG
and the NMG. In the case of the TMG, the BTZ black hole is stable for any
Chern-Simons coupling constant $\mu>0$ by computing (\ref{eql1})
with replacing $M$ by $\mu$, while the stability of the BTZ black hole
is guaranteed for $m^2\ell^2>1/2~(M>\ell)$ by computing two first
order equations (\ref{eql1}) and (\ref{eql2}). In the TMG approach,
the authors~\cite{Birm} have     confirmed the stability by using that a
single massive scalar of $\varphi=z^3h_{zz}$ satisfies the
Klein-Gordon equation in the AdS$_{3}$ background.    On the other
hand, the two left- and right-modes of the NMG [(\ref{psima}) and
(\ref{psimb})] are not orthogonal to each other, implying that the two
are not considered as two independent massive modes propagating on
the BTZ black hole background. This requires a reanalysis of
stability of the BTZ black hole in the conventional black hole approach.
We have obtained two propagating $s$-massive modes $\Phi$ and $\Psi$
by solving the Schr\"odinger equations (\ref{schphi}) and
(\ref{schpsi}) directly, which are surely  not the Schr\"odinger
equations for a massive minimally coupled scalar.

However, the difference between the NMG and the Fierz-Pauli action
is just the presence of the critical point of
$m^2\ell^2=1/2~(c_L=0,c_R=0)$, whose equation is given by the
fourth order linearized equation~(\ref{ceq}). This point exactly
coincides with providing the BTZ black hole solution. Actually,
there is no definite way to solve this equation. In the strict
sense of stability of the BTZ black hole in the NMG, we have
confined ourselves to solving (\ref{ceq}) by using the LCFT
technique in section 5. However, this method is very restrictive
and thus, there is no definite  way to confirm its results at
present. Particularly, the $s$-mode ($k=0$) computation could be
easily done to derive the left- and right logarithmic quasinormal
modes.

At this stage, we would like to comment on how we can apply the
stability condition (\ref{stcon})  of the BTZ black hole in the
NMG to the dual CFT whose central charge is given
by~\cite{bht,LS,AA}
\begin{equation} \label{centralc}
c_{L/R}=\frac{3\ell}{2G}\Big[1-\frac{1}{2m^2 \ell^2}\Big].
\end{equation}
It is well known that the exact agreement is found between the
quasinormal frequencies of the BTZ black hole  and the location of
the poles of the retarded correlation function of the
corresponding perturbations in the CFT~\cite{BSS}. This has
provided  a confirmed test of the AdS$_3$/CFT$_2$ correspondence.
If this correspondence is still valid for the NMG, one
observes  from (\ref{centralc}) that the zero central charge
appears at the critical point, while the unitarity of the CFT$_2$
would demand the bound of $m^2 >1/2\ell^2$ due to $c_{L/R} >0$.
For the non-critical case, we find the stability condition of $M
>\ell$ from (\ref{stcon}), which turns out to be the unitarity
condition for the CFT$_2$. In this case, we expect that the
quasinormal modes obtained  here could be derived from the
location of the poles of the retarded correlation function of the
corresponding perturbations in the CFT$_2$. On the other hand, it
seems  that the quasinomal modes obtained at the critical point of
$M=\ell$ could be found from the retarded correlation function of
the corresponding perturbations in the LCFT as was shown in the
TMG~\cite{Sachs2}. Hence, it is clear  that the stability
condition  of BTZ black hole provides   the unitarity condition of
the dual (L)CFT$_2$.

Finally, we would like to mention that  the linearized  higher dimensional
critical gravities were recently investigated in the AdS
spacetimes~\cite{hcg} but their quasinormal modes are not studied in
the AdS-black hole background.  The non-unitarity issue of log
gravity is not still resolved, indicating that the log gravity
suffers from the ghost problem.

Consequently, we have performed the stability analysis of the BTZ
black hole in the NMG. It seems that the BTZ black hole is stable
against the metric perturbation by computing quasinormal modes and
observing two potentials. However, the stability at the critical
point is not still completely proved because two $s$-modes
[(\ref{1stL}) and (\ref{1stR})] and [(\ref{2ndL}) and (\ref{2ndR})]
are not orthogonal to each other, respectively.

 \vskip 0.5cm

\section*{Acknowledgement}
We would like to thank I. Sachs and B. Tekin for helpful
correspondence. Y.S. Myung and Y.-W. Kim were supported by Basic
Science Research Program through the National Research Foundation
(KRF) of Korea funded by the Ministry of Education, Science and
Technology (2010-0028080). Y.-J. Park was supported by World Class
University program funded by the Ministry of Education, Science and
Technology through the National Research Foundation of Korea(No.
R31-20002). Three of us (Y. S. Myung, T. Moon, and Y.-J. Park) were
also supported by the National Research Foundation of Korea (NRF)
grant funded by the Korea government (MEST) through the Center for
Quantum Spacetime (CQUeST) of Sogang University with grant number
2005-0049409.

\newpage
\section*{Appendix I: LCFT approach to the critical point of NMG}
In this appendix, we will prove the equations of motion
(\ref{2ndeqs}) at the critical point in the BTZ black hole of the NMG
using the LCFT approach. We first note that
\begin{equation}\label{L0y}
L_0 y = \bar{L}_0 y = \frac{1}{2},~~~L_1 y = \bar{L}_1 y =0,
\end{equation}
and
\begin{equation}
L_0 h^{R}_{\mu\nu}=i k h^{R}_{\mu\nu},
 ~~~\bar{L}_0 h^{R}_{\mu\nu} = 0.
\end{equation}
Acting on the R-logarithmic  mode of $h^{R,new}_{\mu\nu}$ with
$L_0(\bar{L}_0)$, we have
\begin{equation}
 L_0 h^{R,new}_{\mu\nu}=i k h^{R,new}_{\mu\nu}+ \frac{1}{2}h^{R}_{\mu\nu},
 ~~~\bar{L}_0 h^{R,new}_{\mu\nu} = \frac{1}{2}h^{R}_{\mu\nu},
\end{equation}
which show  that $h^{R,new}_{\mu\nu}$ is not an eigenstate of $L_0$
or $\bar{L}_0$ as like the new mode at the chiral point in AdS$_3$
of CTMG \cite{GJ}, but it could be an eigenstate of  the subtraction
operator $L_0-\bar{L}_0$.  The two representations of $L_0$ and
$\bar{L}_0$ are now given by the Jordan cell form
\begin{eqnarray}
&& L_0 \left(\begin{array}{c}
     h^{R,new}_{\mu\nu} \\
    h^{R}_{\mu\nu}
  \end{array}\right)
= \left(\begin{array}{cc}
   i k & \frac{1}{2} \\
   0 & i k
  \end{array}\right)
  \left(\begin{array}{c}
     h^{R,new}_{\mu\nu} \\
    h^{R}_{\mu\nu}
  \end{array}\right),  \nonumber\\
&& \bar{L}_0 \left(\begin{array}{c}
     h^{R,new}_{\mu\nu} \\
    h^{R}_{\mu\nu}
  \end{array}\right)
= \left(\begin{array}{cc}
   0 & \frac{1}{2} \\
   0 & 0
  \end{array}\right)
  \left(\begin{array}{c}
     h^{R,new}_{\mu\nu} \\
    h^{R}_{\mu\nu}
  \end{array}\right).
\end{eqnarray}
Thus, it indicates  that $h^{R,new}_{\mu\nu}$ is a L-logarithmic
partner of $h^{R}_{\mu\nu}$. This shows again $h^{R,new}_{\mu\nu}$
(the logarithmic partner of $h^{R}_{\mu\nu}$)  and $h^{R}_{\mu\nu}$
are an eignestate of the subtraction operator $L_0-\bar{L}_0$:
\begin{eqnarray} && \Big[L_0-\bar{L}_0\Big] \left(\begin{array}{c}
     h^{R,new}_{\mu\nu} \\
    h^{R}_{\mu\nu}
  \end{array}\right)
= i k \left(\begin{array}{cc}
   1 & 0 \\
   0 & 1
  \end{array}\right)
  \left(\begin{array}{c}
     h^{R,new}_{\mu\nu} \\
    h^{R}_{\mu\nu}
  \end{array}\right).
\end{eqnarray}
Now, using  the SL$(2,R)$ quadratic Casimir of
$L^2=\frac{1}{2}(L_1L_{-1}+L_{-1}L_1)-L^2_0$, the equation
(\ref{logeom2}) is written by
\begin{equation}
(D^RD^Lh^{R,new})_{\mu\nu}=-(\bar\nabla^2+2)h^{R,new}_{\mu\nu}=
\left[2(L^2+\bar{L}^2)+4\right]h^{R,new}_{\mu\nu}.
\end{equation}
Making use of the following equations
\begin{equation}\label{chwc}
L_1 h^{R,new}_{\mu\nu}=-(2-i k)e^u\tanh(\rho)h^{R,new}_{\mu\nu},
~~~\bar{L}_1 h^{R,new}_{\mu\nu} = 0,
\end{equation}
we arrive at
\begin{equation} \label{ddnew}
(D^RD^Lh^{R,new})_{\mu\nu}=-(\bar\nabla^2+2)h^{R,new}_{\mu\nu}=-2h^R_{\mu\nu}.
\end{equation}
Operating $(\bar{\nabla}^2-2\Lambda)$ on (\ref{ddnew}) with
$\Lambda=-1$ and using (\ref{seq1}) on the R-moving mode
$h^R_{\mu\nu}$, this leads to (\ref{ceq}): $(D^RD^L)h^{R,new}\not=0,
(D^RD^L)^2h^{R,new}=0$. We remind the reader  that $\bar{L}_1
h^{R,new}_{\mu\nu}(=0)$ in (\ref{chwc}) is the ``chiral" highest
weight condition, but not $L_1 h^{R,new}_{\mu\nu}\neq 0$ as shown in
(\ref{chwc}). This differs clearly from the CTMG at the chiral point
of the AdS$_3$ background~\cite{GJ},  where one should impose the
``highest weight" conditions of both $L_1
h^{R,new}_{\mu\nu}=\bar{L}_1 h^{R,new}_{\mu\nu}(=0)$~\cite{Myung}.

On the other hand, for the new L-logarithmic mode (\ref{logeom1}),
we  follow the same steps as did in the new R-moving mode. In
addition to (\ref{L0y}), we have
\begin{equation}
L_0 h^{L}_{\mu\nu}= 0,
 ~~~\bar{L}_0 h^{L}_{\mu\nu} = -i k h^{L}_{\mu\nu} .
\end{equation}
Therefore, the new L-logarithmic  mode of $h^{L,new}_{\mu\nu}$
satisfies
\begin{equation}
 L_0 h^{L,new}_{\mu\nu}=\frac{1}{2}h^{L}_{\mu\nu},
 ~~~\bar{L}_0 h^{L,new}_{\mu\nu} = - i k h^{L,new}_{\mu\nu}+\frac{1}{2}h^{L}_{\mu\nu}.
\end{equation}
Similarly, the two  representations of $L_0$ and $\bar{L}_0$ take
the compact matrix forms
\begin{eqnarray}
&& L_0 \left(\begin{array}{c}
     h^{L,new}_{\mu\nu} \\
    h^{L}_{\mu\nu}
  \end{array}\right)
= \left(\begin{array}{cc}
   0 & \frac{1}{2} \\
   0 & 0
  \end{array}\right)
  \left(\begin{array}{c}
     h^{L,new}_{\mu\nu} \\
    h^{L}_{\mu\nu}
  \end{array}\right),  \nonumber\\
&& \bar{L}_0 \left(\begin{array}{c}
     h^{L,new}_{\mu\nu} \\
    h^{L}_{\mu\nu}
  \end{array}\right)
= \left(\begin{array}{cc}
   -ik & \frac{1}{2} \\
   0 & -ik
  \end{array}\right)
  \left(\begin{array}{c}
     h^{L,new}_{\mu\nu} \\
    h^{L}_{\mu\nu}
  \end{array}\right).
\end{eqnarray}
This shows again $h^{L,new}_{\mu\nu}$ (the logarithmic partner of
$h^{L}_{\mu\nu}$)  and $h^{L}_{\mu\nu}$ are an eignestate of the
subtraction operator $L_0-\bar{L}_0$: \begin{eqnarray} &&
\Big[L_0-\bar{L}_0\Big] \left(\begin{array}{c}
     h^{L,new}_{\mu\nu} \\
    h^{L}_{\mu\nu}
  \end{array}\right)
= i k \left(\begin{array}{cc}
   1 & 0 \\
   0 & 1
  \end{array}\right)
  \left(\begin{array}{c}
     h^{L,new}_{\mu\nu} \\
    h^{L}_{\mu\nu}
  \end{array}\right).
\end{eqnarray}
Now, making use of the following equations
\begin{equation}
L_1 h^{L,new}_{\mu\nu}=0, ~~~\bar{L}_1 h^{L,new}_{\mu\nu} = -(2+i
k)e^v \tanh(\rho) y(\tau,\rho) h^{L}_{\mu\nu},
\end{equation}
where $L_1 h^{L,new}_{\mu\nu}(=0)$ is the ``anti-chiral" highest
weight condition, but not for $\bar{L}_1 h^{L,new}_{\mu\nu}\neq
0$, we finally arrive at
\begin{equation}
(D^RD^Lh^{L,new})_{\mu\nu}=-(\bar\nabla^2+2)h^{L,new}_{\mu\nu}=-2h^L_{\mu\nu},
\end{equation}
which  clearly confirms (\ref{2ndeqs}).

\newpage
\section*{Appendix II: Full perturbation analysis in BTZ black hole background}
Considering  the full $h_{\mu\nu}$ components given by
\begin{eqnarray}
h_{\mu\nu}=\left(
\begin{array}{ccc}
H_{tt}(r) & H_{tr}(r) & H_{t\phi}(r) \cr H_{rt}(r) & H_{rr}(r) &
H_{r\phi}(r) \cr H_{\phi t}(r) & H_{\phi r}(r) & H_{\phi\phi}(r)
\end{array}
\right) e^{\omega_h t}e^{ik\phi} \,,
\end{eqnarray}
the tensor  perturbation equation of  $\bar{\nabla}^2_{\rm BTZ}
h_{\mu\nu}-\Big(\frac{5\Lambda}{2}+m^2\Big)h_{\mu\nu}=0$ lead to\\
\begin{eqnarray}
(t,t);&&-2r^2({\cal M}+\Lambda r^2)^2 H_{tt}''-2r({\cal
M}^2-\Lambda^2 r^4)H_{tt}'-\left\{(2{\cal M}k^2+2\Lambda
m^2r^2)-\Lambda r^2(3{\cal M}-\Lambda
r^2)\right.\nonumber\\
&&\hspace{-1em}\left.-2r^2\omega_h^2+2m^2r^2({\cal M}+\Lambda
r^2)\right\}H_{tt}+ 4\Lambda^2 r^4({\cal M}+\Lambda r^2)^2H_{rr}
-8\Lambda\omega_h r^3({\cal
M}+\Lambda r^2)H_{tr}=0\nonumber\\
&&\nonumber\\
(t,r);&&2r^3({\cal M}+\Lambda r^2)^3H_{tr}''+2r^2({\cal M}+\Lambda
r^2)^2({\cal M}+3\Lambda r^2)H_{tr}'+r({\cal M}+\Lambda
r^2)\left\{-2{\cal M}^2+\Lambda r^4\times\right.\nonumber\\
&&\left.(2m^2 -5\Lambda)+{\cal M}r^2(\Lambda+2m^2)+2k^2({\cal
M}+\Lambda r^2)-2r^2\omega_h^2\right\}H_{tr}+4ik({\cal M}+\Lambda
r^2)^2H_{t\phi}\nonumber\\
&&+4\Lambda\omega_h r^4H_{tt}+4\Lambda\omega_h r^4({\cal
M}+\Lambda r^2)^2H_{rr}=0\nonumber\\
&&\nonumber\\
(t,\phi);&&2r^2({\cal M}+\Lambda r^2)^2H_{t\phi}''-2r({\cal
M}+\Lambda r^2)^2H_{t\phi}'+\left\{({\cal M}+\Lambda
r^2)(2k^2+(\Lambda+2m^2)r^2)\right.\nonumber\\
&&\left.-2r^2\omega_h^2\right\}H_{t\phi}+4ikr({\cal M}+\Lambda
r^2)^2H_{tr}+4\Lambda\omega_h r^3({\cal
M}+\Lambda r^2)H_{r\phi}=0\nonumber\\
&&\nonumber\\
(r,r);&&-2r^4({\cal M}+\Lambda r^2)^4H_{rr}''-2r^3({\cal M}+\Lambda
r^2)^3({\cal M}+7\Lambda r^2)H_{rr}'-r^2({\cal M}+\Lambda
r^2)^2\left\{-4{\cal
M}^2+\right.\nonumber\\
&&\left.{\cal M}r^2(2m^2+5\Lambda)+\Lambda
r^4(2m^2+13\Lambda)+2k^2({\cal M}+\Lambda r^2)-2\omega_h^2
r^2\right\}H_{rr}+4\Lambda^2 r^6H_{tt}+\nonumber\\
&&-8\Lambda\omega_h r^5 ({\cal M}+\Lambda r^2)H_{tr}+4({\cal
M}+\Lambda r^2)^3H_{\phi\phi}-8ikr({\cal
M}+\Lambda r^2)^3H_{r\phi}=0\nonumber\\
&&\nonumber\\
(r,\phi);&&2r^3({\cal M}+\Lambda r^2)^3H_{r\phi}''-2r^2({\cal
M}+\Lambda r^2)^2({\cal M}-3\Lambda r^2)H_{r\phi}+r({\cal M}+\Lambda
r^2)\left\{({\cal M}+\Lambda r^2)\right.\nonumber\\
&&\left.\times(2k^2-6{\cal M})+r^2({\cal M}+\Lambda
r^2)(2m^2-5\Lambda)-2\omega_h^2 r^2\right\}H_{r\phi}+4ikr^2({\cal
M}+\Lambda
r^2)^3H_{rr}\nonumber\\
&&+4\Lambda\omega_h r^4H_{t\phi}+4ik({\cal M}+\Lambda
r^2)^2H_{\phi\phi}=0\nonumber\\
&&\nonumber\\
(\phi,\phi);&&2r^2({\cal M}+\Lambda r^2)^2H_{\phi\phi}''-2r({\cal
M}+\Lambda r^2)(3{\cal M}+\Lambda r^2)H_{\phi\phi}'+\left\{({\cal
M}+\Lambda r^2) (2k^2+4{\cal
M}+\right.\nonumber\\
&&\left.2m^2r^2+\Lambda r^2)-2\omega_h^2r^2\right\}H_{\phi\phi}
-4r^2({\cal M}+\Lambda r^2)^3H_{rr}+8ikr({\cal M}+\Lambda
r^2)^2H_{r\phi}=0\nonumber.
\end{eqnarray}
Also the TT gauge condition of $\bar{\nabla}^{\mu}h_{\mu\nu}=0$ and
$h=0$ are given by
\begin{eqnarray}
t;&&r^2({\cal M}+\Lambda r^2)^2H_{tr}'+r({\cal M}+\Lambda r^2)({\cal
M}+3\Lambda r^2)H_{tr}-i k({\cal M}+\Lambda
r^2)H_{t\phi}+r^2\omega_h
H_{tt}=0\nonumber\\
r;&&r^3({\cal M}+\Lambda r^2)^3H_{rr}'+r^2({\cal M}+\Lambda
r^2)^2({\cal M}+4\Lambda r^2)H_{rr}+\Lambda r^4H_{tt}+({\cal
M}+\Lambda r^2)^2H_{\phi\phi}\nonumber\\
&&-\omega_h r^3({\cal M}+\Lambda
r^2)H_{tr}-i k r({\cal M}+\Lambda r^2)^2H_{r\phi}=0\nonumber\\
\phi;&&\hspace{-1em}-r^2({\cal M}+\Lambda r^2)^2H_{r\phi}'-r({\cal
M}+\Lambda r^2)({\cal M}+3\Lambda r^2)H_{r\phi}+i k ({\cal
M}+\Lambda
r^2)H_{\phi\phi}+\omega_h r^2H_{t\phi}=0\nonumber\\
&&\nonumber\\
&&r^2H_{tt}-r^2({\cal M}+\Lambda r^2)^2H_{rr}+({\cal M}+\Lambda
r^2)H_{\phi\phi}=0\nonumber
\end{eqnarray}
 For type I(odd)  metric ansatz (\ref{oddp}), i.e., $H_{tt}=H_{tr}=H_{rr}=H_{\phi\phi}=0$,
  $(t,r)$ equation becomes
\begin{eqnarray}
4ik({\cal M}+\Lambda r^2)^2H_{t\phi}=0,\nonumber
\end{eqnarray}
and the solution is $H_{t\phi}=0$ for $k\neq0$. However, this
corresponds to  the null odd-solution because in this case, we
obtain $H_{r\phi}=0$ from $(t,\phi)$ equation.  Therefore, the
$s$-mode ($k=0$) solution is only admitted for type I metric ansatz
(\ref{oddp}).

On the other hand, when focusing on $(t,\phi)$
equation and considering type II (even) metric ansatz (\ref{evenp})
with $H_{t\phi}=H_{r\phi}=0$, equation $(t,\phi)$ reduces to
\begin{eqnarray}
4ikr({\cal M}+\Lambda r^2)^2H_{tr}=0.\nonumber
\end{eqnarray}
Unless $k=0$, we obtain  $H_{tr}=0$. In this case, $H_{tt}=0$ is
found from $t$ component equation of
$\bar{\nabla}^{\mu}h_{\mu\nu}=0$. Furthermore, we have  $H_{rr}=0$
and  $H_{\phi\phi}=0$ from  $(t,t)$ and  $(r,r)$ equations. This
corresponds to the null even-solution.  So we also have to restrict to
the $s$-mode case. In other words, the $s$-mode case leads to type I
and II metric splitting for obtaining two  modes of  a massive
graviton.

\end{document}